\documentclass[preprint]{vgtc}               % preprint
\graphicspath{{figures/}{pictures/}{images/}{./}} % where to search for the images

\usepackage{times}                     % we use Times as the main font
\usepackage{amsmath}

\usepackage{mathptmx}                  % use matching math font

\onlineid{1054}

\vgtccategory{technique or algorithm}

\vgtcinsertpkg

\ieeedoi{10.1109/VIS55277.2024.00042}

\usepackage{pgfplots}
\pgfplotsset{compat=newest}

\usepackage{subcaption}

\usepackage[super]{nth}

\usepackage{booktabs}
\usepackage{multirow}
\usepackage{tabularray}

\usepackage{fmtcount}

\usepackage{adjustbox}

\usepackage{cancel}
\usepackage{xcolor}

\title{Accelerating Transfer Function Update for\\Distance Map based Volume Rendering}

\author{Michael Rauter\thanks{e-mail: michael.rauter@fhwn.ac.at}\\
	\parbox{2.05in}{\scriptsize \centering Competence Center for Preclinical Imaging and Biomedical Engineering, Faculty of Health, University of Applied Sciences Wiener Neustadt, Austria}
\and Lukas Zimmermann\\
		\parbox{2.05in}{\scriptsize \centering Department of Radiation Oncology, \\Medical University of Vienna, Austria}
\and Markus Zeilinger\\
     \parbox{2.05in}{\scriptsize \centering Competence Center for Preclinical Imaging and Biomedical Engineering, Faculty of Health, University of Applied Sciences Wiener Neustadt, Austria}
}

\teaser{
	\vspace{-1mm}
  \centering
  \includegraphics[width=\linewidth]{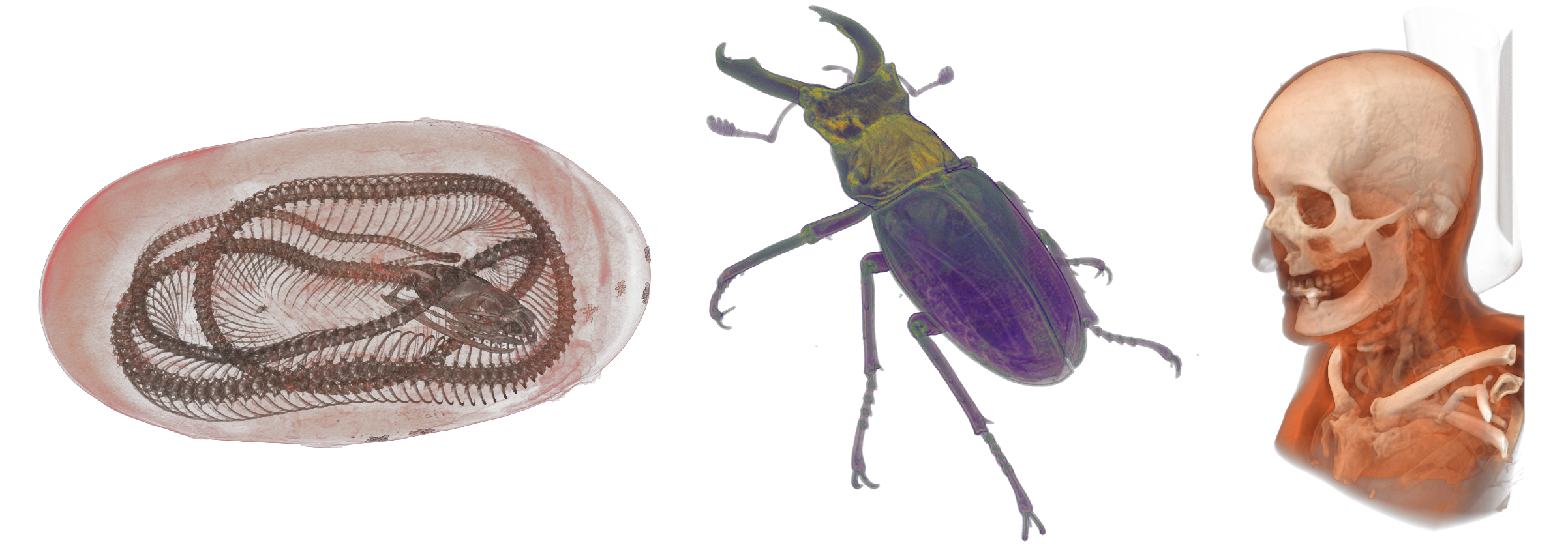}
  \caption{Volume renderings for kingsnake~\cite{kingsnake}, stag beetle~\cite{stag_beetle}, and manix~\cite{osirix} datasets applying different transfer functions}
  \label{fig:teaser}
}

\abstract{Direct volume rendering using ray-casting is widely used in practice. By using GPUs and applying acceleration techniques as empty space skipping, high frame rates are possible on modern hardware. This enables performance-critical use-cases such as virtual reality volume rendering.
The currently fastest known technique uses volumetric distance maps to skip empty sections of the volume during ray-casting but requires the distance map to be updated per transfer function change.
In this paper, we demonstrate a technique for subdividing the volume intensity range into partitions and deriving what we call partitioned distance maps. These can be used to accelerate the distance map computation for a newly changed transfer function by a factor up to 30.
This allows the currently fastest known empty space skipping approach to be used while maintaining high frame rates even when the transfer function is changed frequently.
}


\begin{document}

%% The ``\maketitle'' command must be the first command after the
%% ``\begin{document}'' command. It prepares and prints the title block.

%% the only exception to this rule is the \firstsection command
\firstsection{Introduction}

\maketitle

%% \section{Introduction} %for journal use above \firstsection{..} instead
Direct volume rendering is an important technique to get insight into a volume dataset.
Direct volume rendering has the advantage of not relying on volume segmentations while producing high quality renders.
The main technique in direct volume rendering is volume ray casting where a ray is cast from the camera eye point to every pixel in the target render image~\cite{levoy_1988, krueger_2003}.
Discrete steps are taken along the ray covering the volume's bounding geometry. At each position the corresponding volume's voxel values are sampled and interpolated.
Those sampled voxel values are then mapped using a transfer function (TF) to rendered color and opacity values as described by Kniss et al.~\cite{kniss_2002}. This can be done by either using a 1D mapping from intensities, or a 2D mapping from intensities and gradient magnitude.
Voxels along a single ray's path are blended together for the final projected pixel color in the rendered image.

The high computational demands of volume rendering pose a challenge but with the growing computing performance of graphics processing units (GPUs) in recent years and optimized acceleration techniques volume rendering with high interactive frame-rates and high rendering resolutions became a reality. Even especially demanding use cases as volume rendering in virtual reality are possible~\cite{fons_2018, elbeheiry_2019, kalarat_2019, faludi_2019_complete, scholl_2019, waschk_2020, nysjo_2020, faludi_2021, gao_2022}.

Two important acceleration techniques are early ray termination and empty space skipping. Early ray termination stops the ray stepping iteration for a given pixel as soon as the aggregated opacity exceeds a certain threshold. Empty space skipping allows to step over voxel positions along the ray that do not contribute to the output pixel's color. This is done by using an acceleration data structure providing information about how much space can be skipped\cite{engel_2006}.

Our paper demonstrates an acceleration technique for a certain part of distance map based empty space skipping as described by Deakin and Knackstedt~\cite{deakin_2019}. We show how the acceleration data structure (the distance map) can be recomputed faster which is required in case the TF is altered. This is achieved by subdividing the whole volume's intensity range into partitions. We derive individual distance maps from the respective intensity partitions. Given a newly changed TF, these partitioned distance maps can be efficiently combined into an updated distance map.

\section{Related work}
Empty space skipping (ESS) methods can be classified into object-order and image-order ESS methods~\cite{lacroute_1994}. Here, we focus on the latter class.

Krüger and Westermann use an occupancy map encoding disjoint blocks of $8^3$ voxels storing the blocks' minimum and maximum intensity values with non-zero transparency contributions~\cite{krueger_2003}. This basically amounts encoding a single level of an min/max octree, in contrast to standard min/max octree implementations~\cite{lacroute_1994, wald_2007}. Alternatively, the acceleration structure can be an octree of binary values encoding the occupancy of its nodes as employed by Levoy~\cite{levoy_1990}.

SparseLeap is a hybrid object- and image-order ESS method proposed by Hadwiger et al.~\cite{hadwiger_2018}.
An occupancy histogram tree is used to extract the dataset's bounding geometry. Per-pixel linked lists of skippable ray segments are created to efficiently skip empty space when rendering the volume.
A drawback is the requirement of recomputing the histogram tree and the bounding geometry during TF changes.

A different approach uses distance maps storing the smallest distance to the nearest non-transparent block of voxels or rectilinear grids as proposed by Sramek and Kaufman~\cite{sramek_2000}.

Deakin and Knackstedt~\cite{deakin_2019} generate a reduced occupancy map and the corresponding Chebyshev distance map encoding the distance to the nearest non-transparent block of voxels utilizing GLSL compute shaders on the GPU. In terms of rendering time, their volume renderer outperforms other approaches, e.g., depending on the dataset used, they report runtimes on average twice as fast as octree-based methods and SparseLeap. Recomputing the distance map on a TF change is a limiting performance factor in their approach.

Faludi et al.~\cite{faludi_2021} use an octree of nodes storing a bitfield, where each bit inside the bitfield encodes the occupancy of an intensity subrange. %inside a subrange of the full intensity range.
Similarly, a bitfield represenation of the currently selected TF is derived. Testing if an octant only contains transparent voxels according to the TF and thus may be skipped, can be determined with a binary operation between its bitfield and the bitfield of the TF. Changing the TF just requires recomputing the TF bitfield and does not affect performance noticeably. They report ray marching times 35\% slower than Deakin's approach on average.

\section{Method}
We introduce an optimization technique for updating the distance map in volume rendering using distance map based ESS. Our approach allows computing the distance map in less time than before by utilizing precomputed individual distance maps encoding voxel occupancies inside specific intensity subranges. The update step is an aggregation step combining several of those distance maps to produce the final distance map used for ESS.
The TF dictates which distance maps require merging.

\subsection{Intensity-Partitioned Occupancy/Distance Maps} \label{sec:POMs_and_PDMs_explained}
Our approach is inspired by the idea of subdividing the intensity range similar to how it was done by Faludi et al.~\cite{faludi_2021}. We also split the intensity range into multiple subranges but instead of storing the occupancy information in an octree of bitfields we store it in individual distance maps for the different intensity partitions.
The total range of intensity values used in the volume dataset is subdivided into $n$ intensity partitions. The partitions can be uniformly distributed over the whole intensity range, but this is not a necessary requirement. For example, it can be beneficial to use a separate partition for the minimum intensity value $\rho_{min}$ encountered inside the volume and distribute the other intensity values to the remaining partitions.
A single partition $p$ is defined by the lower and upper border of its intensity range denoted as integer indices $\rho_{min}^p$ and $\rho_{max}^p$.

We compute $n$ block-reduced occupancy maps for the respective subranges which we call partitioned occupany maps (POMs) from now on.
A block of voxels is considered occupied if it contains at least one voxel with an intensity value from within the intensity subrange $\left[\rho_{min}^p,\rho_{max}^p\right]$.
Block-reduced means that the original volume dataset is subdivided into non-overlapping blocks of voxels of size $b^3$.
We set $b$ to 4 which was also used as block size by Deakin and Knackstedt~\cite{deakin_2019}.
It is important to point out that the POMs do only depend on the volume's voxel intensity values and not on the TF used for rendering.

Next, we compute $n$ partitioned distance maps (PDMs) from the $n$ individual POMs. The PDMs encode the Chebyshev distance to non-empty voxels in the respective intensity subrange.

Both the POMs as well as the PDMs are in block-reduced size and their computation can be done in a preprocessing step (this resembles a one-time initialization step which only needs to be done once when a volume is initially loaded).

Whenever the TF is loaded or updated, merging the relevant PDMs into a final single distance map $D'$ is done (the TF dictates the map selection required for merging). $D'$ is the only entity directly used for ESS in the renderer. $D'$ is assembled in two steps:\\
Step 1: Determine all intensity partitions containing at least one value getting mapped by the TF to a non-zero alpha value.\\
Step 2: Compute the element-wise minimum of the corresponding PDMs. This resembles the combined distance map $D'$ in~\autoref{equ:distancemap_combination}.
\begin{equation}
D^\prime(x,y,z) =  {\overset{p \in S}{min}}(PDM_p(x,y,z))
\label{equ:distancemap_combination}
\end{equation}
$S$ is the set of indices $p$ of partitions determined in step 1, $x,y,z$ are the integer coordinates of a block.

\begin{figure}[!h]
	\centering
		\includegraphics[width=0.48\textwidth]{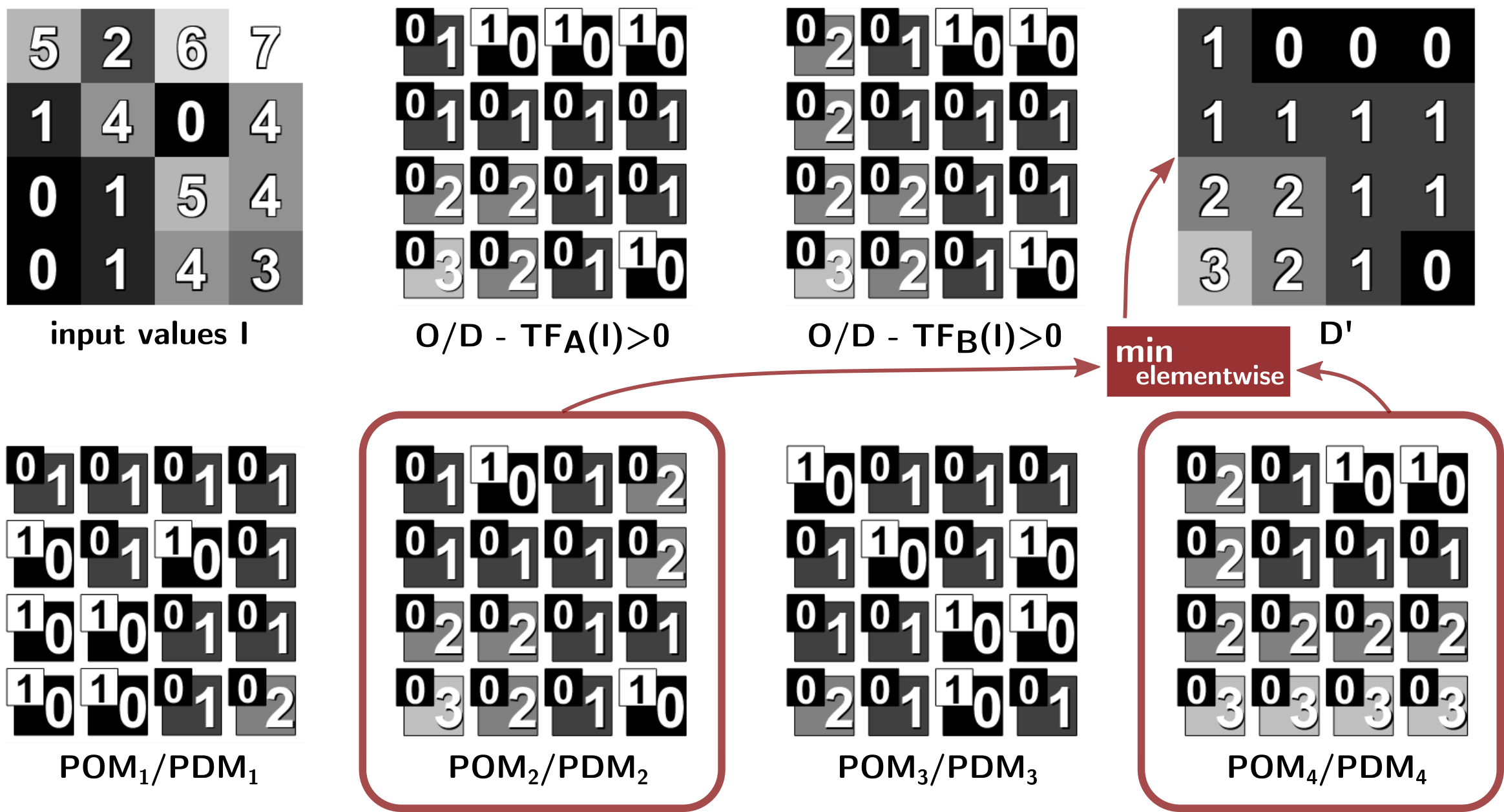}
	\caption{
Our method demonstrated for a simplified 2D case (without block-reducing): \ordinalnum{1} row from left to right: input intensity map I, occupancy and corresponding distance map (small squares contain occupancy values) after applying TF\textsubscript{A} resp. TF\textsubscript{B},\vspace{0.5mm} $D'$ derived from the PDMs ($D' = D'_{\mathrm{TF}_A} = D'_{\mathrm{TF}_B}$ in this example),	\ordinalnum{2} row: the POMs and PDMs using partitions with intensity ranges $[0,1]$, $[2,3]$, $[4,5]$, and $[6,7]$; TFs' intensities to alpha mappings ($\cancel{0}\ldots$ non-zero alpha):
$\mathrm{TF}_A(0,\ldots,7) = \left\{0,0,\cancel{0},\cancel{0},0,0,\cancel{0},\cancel{0}\right\}$, $\mathrm{TF}_B(0,\ldots,7) = \left\{0,0,0,\cancel{0},0,0,\cancel{0},\cancel{0}\right\}$ and thus, $S = \left\{2,4\right\}$ for both.
	}
	\label{fig:method_example_draft}
\end{figure}

See~\autoref{fig:method_example_draft} for a simplified 2D example of our technique.
In general, $D^\prime$ may differ from the standard non-partitioned distance map ($D$). This happens if different intensity values from the same intensity partition get mapped to both zero and non-zero alpha values by the TF. The values of $D^\prime$ are always an element-wise lower limit of the values in $D$ (e.g. $D^\prime \leq D$). As a consequence ESS run-time performance may be worse because less voxels will be skipped. It can never happen that more voxels than allowed are skipped making $D^\prime$ a reasonable approximation and thus replacement of $D$. Again, refer to~\autoref{fig:method_example_draft} and compare $D'$ to $D$ when applying TF\textsubscript{A} ($D' = D$) and when applying TF\textsubscript{B} ($D' \leq D$).
$D^\prime$ is - as already stated - the only component that needs to be recomputed when a TF update occurs. Taking the element-wise minimum of the distance values from the PDMs is computationally less expensive than computing the actual distance transform algorithm which makes our approach more efficient.

\begin{table*}[!t]
	\caption{Comparison of time required to compute the acceleration structure used for ESS of the original approach~\cite{deakin_2019} compared to our approach with different numbers of intensity partitions. For each dataset, we report timings for the different TFs used (TFs are those from~\autoref{fig:transfer_functions_used}).}
\resizebox{17.7cm}{!}{

\begin{tblr}{
	colspec={llllcccccc},
	%vline{3}={2}{4}{solid}
	%vline{3}={1}{-}{abovepos=-20}
	vline{3,4} = {2-22}{solid}
}
    \hline
    dataset &
		size &
		computation type &
		\begin{tabular}[c]{@{}l@{}}transfer\\ function\end{tabular} &
		\begin{tabular}[c]{@{}l@{}}distance\\ map~\cite{deakin_2019}\end{tabular} &
		\begin{tabular}[c]{@{}l@{}}PDMs\\ (16 partitions)\end{tabular} &
		\begin{tabular}[c]{@{}l@{}}PDMs\\ (32 partitions)\end{tabular} &
		\begin{tabular}[c]{@{}l@{}}PDMs\\ (64 partitions)\end{tabular} &
		\begin{tabular}[c]{@{}l@{}}PDMs\\ (128 partitions)\end{tabular} &
		\begin{tabular}[c]{@{}l@{}}PDMs\\ (256 partitions)\end{tabular}	\\
    \hline
    \SetCell[r=7]{} manix &  \SetCell[r=7]{} 512 × 512 × 460 & one-time initialization {[}ms{]} & TF1-TF6 & - & 58.1 & 115.3 & 227.9 & 454.4 & 905.2\\ \cline{3-10}
    & & \SetCell[r=6]{} update time on TF change {[}ms{]} & TF1 & 3.50 & 0.20 & 0.27 & 0.42 & 0.69 & 1.30\\ \cline{4-10}
    & & & TF2 & 2.55 & 0.20 & 0.26 & 0.44 & 0.70 & 1.27 \\ \cline{4-10}
    & & & TF3 & 4.31 & 0.18 & 0.19 & 0.28 & 0.44 & 0.69 \\ \cline{4-10}
    & & & TF4 & 4.03 & 0.17 & 0.22 & 0.28 & 0.41 & 0.71 \\ \cline{4-10}
    & & & TF5 & 3.85 & 0.20 & 0.26 & 0.38 & 0.58 & 0.98 \\ \cline{4-10}
    & & & TF6 & 3.66 & 0.18 & 0.20 & 0.26 & 0.44 & 0.71 \\ \cline{1-10}
    \SetCell[r=7]{} stag beetle &  \SetCell[r=7]{} 832 × 832 × 494 & one-time initialization {[}ms{]} & TF1-TF6 & - & 146.4 & 299.9 & 608.7 & 1233.0 & 2491.8\\ \cline{3-10}
    & & \SetCell[r=6]{} update time on TF change {[}ms{]} & TF1 & 10.10 & 0.41 & 0.55 & 0.87 & 1.40 & 2.58\\ \cline{4-10}
    & & & TF2 & 0.40 & 0.41 & 0.58 & 0.85 & 1.38 & 2.59 \\ \cline{4-10}
    & & & TF3 & 12.23 & 0.31 & 0.39 & 0.58 & 0.81 & 1.40 \\ \cline{4-10}
    & & & TF4 & 11.30 & 0.33 & 0.40 & 0.58 & 0.80 & 1.42 \\ \cline{4-10}
    & & & TF5 & 11.13 & 0.45 & 0.51 & 0.74 & 1.14 & 2.03 \\ \cline{4-10}
    & & & TF6 & 11.12 & 0.33 & 0.40 & 0.57 & 0.85 & 1.42 \\ \cline{1-10}
    \SetCell[r=7]{} kingsnake &  \SetCell[r=7]{} 1024 × 1024 × 795 & one-time initialization {[}ms{]} & TF1-TF6 & - & 228.2 & 458.2 & 913.7 & 1825.9 & 3659.2\\ \cline{3-10}
    & & \SetCell[r=6]{} update time on TF change {[}ms{]} & TF1 & 9.63 & 0.80 & 1.25 & 1.97 & 3.58 & 6.71\\ \cline{4-10}
    & & & TF2 & 7.28 & 0.80 & 1.25 & 1.99 & 3.56 & 6.70 \\ \cline{4-10}
    & & & TF3 & 21.40 & 0.63 & 0.77 & 1.23 & 1.94 & 3.58 \\ \cline{4-10}
    & & & TF4 & 14.59 & 0.63 & 0.76 & 1.23 & 1.94 & 3.59 \\ \cline{4-10}
    & & & TF5 & 15.28 & 0.79 & 1.08 & 1.68 & 2.86 & 5.39 \\ \cline{4-10}
    & & & TF6 & 9.76 & 0.66 & 0.79 & 1.26 & 2.09 & 3.60 \\ \cline{1-10}
    %\hline
\end{tblr}

}

	\label{table:timings_distance_maps_computation}
\end{table*}

\subsection{Implementation}
We implemented our method in a custom direct ray-casting volume rendering shader using OpenGL/GLSL with C++. Rendered images from our renderer can be seen in~\autoref{fig:teaser}. The computation of the POMs and PDMs as well as computing $D^\prime$ was implemented with GLSL Compute Shaders. POMs and PDMs use an \textit{uint8} data type for performance and memory consumption reasons, distances higher than the maximum unsigned 8-bit integer value are clamped. Combining the PDMs in the compute shader is done in several passes, aggregating distances by binding only distance maps as input textures that have a contribution to $D^\prime$ according to the TF. In every pass up to a maximum of 6 PDMs are bound and combined (2 additional textures being used for flip-flopping input (result from previous passes) and output texture). The number of necessary passes depend on the number of PDMs that need to be combined derived from the non-zero alpha mappings in the TF.

\section{Results}
To demonstrate the benefits of our approach we conducted a series of experiments to show the performance gain for the distance map update on a TF change and the implications to the frame rendering performance. A video showing volume rendering results with frame rendering performance is provided as supplementary material.

\subsection{Benchmark Configuration}
For our performance tests, we use a \textit{Windows11} workstation equipped with an \textit{Intel Core i7-13700KF} CPU running at 3.4 GHz, 32 GB RAM and a \textit{NVIDIA RTX 4090} GPU. The timing measurements for the computation times of the distance maps generation and update depend on the size of the volume dataset, the TF used and the intensity partition configuration chosen (we benchmarked 16, 32, 64, 128, and 256 partitions). The timing measurements for frame rendering depend on the viewport size, the viewport coverage by the volume, the TF used and the chosen rendering options/parameters (e.g., if early ray termination or ESS is used). Viewport size was chosen as $2160^2$, we made the rendering of the volume cover the vertical viewport extent, early ray termination (if used) was set to terminate the ray at $0.98$ of the maximum possible alpha value. For the frame rendering benchmark, the camera was rotated around the vertical axis going through the center of the volume datasets completing two full rotations over the time period of 10 seconds. With this setup, we tried to reproduce the setup chosen by Faludi et al.~\cite{faludi_2021}.
\subsection{Performance Evaluation}
\begin{figure}[!tb]
	\centering % avoid the use of \begin{center}...\end{center} and use \centering instead (more compact)
\begin{tikzpicture}[
		node font=\footnotesize,
		farbe/.style={draw=#1!80!black,fill=#1!20},
		/pgfplots/every axis/.style={
				ybar stacked,
				xtick=data,
				xmin=0.25,
				xmax=7.75,
				ymin=0,
				ymax=13.5,
				ytick={0,2,...,14},
				bar width=4.5pt,
				legend style={nodes={scale=0.8, transform shape}},
				legend image code/.code={%
											\draw[#1, draw=none] (0cm,-0.1cm) rectangle (0.2cm,0.033cm);
				},
			},
			]

		\definecolor{barcolor1_ert}{RGB}{210, 90, 100}
		\definecolor{barcolor2_ert}{RGB}{21, 15, 110}
		\definecolor{barcolor3_ert}{RGB}{25, 109, 60}
		\definecolor{barcolor_acceleration_structure_update}{RGB}{200, 180, 160}

		\pgfplotsset{
					height=3.0cm, width=7cm,
					scale only axis, 
					legend style={draw=none,legend cell align=left,}
		}
		% Barplots
		\begin{axis}[
				name=main_axis,
				bar shift=-3pt,
				bar width=9pt,
				xticklabels={No ESS, Block ESS , Distance ESS ,},     
				ylabel={Frame Time [ms]},
				x tick label style={align=left, rotate=45, anchor=east, font=\small},
				xtick style={draw=none},
				y tick label style={
					/pgf/number format/fixed,
					/pgf/number format/fixed zerofill,
					/pgf/number format/precision=1
        },
				minor y tick num={1},
				ylabel={Frame Time [ms]},			
				ymajorgrids=true,
				yminorgrids=true,
				major grid style={dotted,black},
				minor grid style={dotted,black},
			] 

			\addplot[farbe=barcolor1_ert] coordinates 
				{(1, 4.72) (2, 1.98) (3, 1.34)};
			\label{pgfplots:reference_methods}
			\addplot[farbe=barcolor_acceleration_structure_update] coordinates 
				{(1, 0.0) (2, 3.17) (3, 11.12)};
				
		\end{axis}
		
		\begin{axis}[bar shift=-3pt,
				xticklabels={PDM ESS\\[-3pt]\adjustbox{lap =.15\width}{\small{(32 partitions)}}, PDM ESS\\[-3pt]\adjustbox{lap =.15\width}{\small{(64 partitions)}}, PDM ESS\\[-3pt]\adjustbox{lap =.15\width}{\small{(128 partitions)}}, PDM ESS\\[-3pt]\adjustbox{lap =.15\width}{\small{(256 partitions)}},},
				%x tick label style={text width=2.5cm, rotate=45, anchor=east},
				x tick label style={align=right, rotate=45, anchor=east, xshift=0.1cm, yshift=-0.3cm, font=\small},
				xtick style={draw=none},
				%y tick label style={draw=none},
				yticklabel=\empty,
				%y tick label style={
					%/pgf/number format/fixed,
					%/pgf/number format/fixed zerofill,
					%/pgf/number format/precision=1
        %},
				%minor y tick num={1},
				%ylabel={Frame Time [ms]},			
				%ymajorgrids=true,
				%yminorgrids=true,
				%major grid style={dotted,black},
				%minor grid style={dotted,black},
		]
				\addplot+[farbe=barcolor2_ert] coordinates
				{(4, 4.72) (5, 1.34) (6, 1.34) (7, 1.34)};
				\label{pgfplots:uniform_partitions}
				\addplot+[farbe=barcolor_acceleration_structure_update] coordinates
				{(4, 0.40) (5, 0.57) (6, 0.85) (7, 1.42)};
		\end{axis}
		
		\begin{axis}[bar shift=3pt,hide axis]
				\addplot+[farbe=barcolor3_ert] coordinates
				{(4, 1.34) (5, 1.34) (6, 1.34) (7, 1.34)};
				\label{pgfplots:special_partition}
				\addplot+[farbe=barcolor_acceleration_structure_update] coordinates
				{(4, 0.40) (5, 0.57) (6, 0.85) (7, 1.42)};
				\label{pgfplots:acceleration_struct_update}
		\end{axis}
		
		\node[draw,fill=white,inner sep=0pt,below left=0.5em] at (main_axis.north east) {
		\small
		\begin{tabular}{llllll}
		\ref{pgfplots:reference_methods} comparison methods\\
		\ref{pgfplots:uniform_partitions} uniform partitions\\
		\ref{pgfplots:special_partition} special partition for $\rho_{min}$\\
		\ref{pgfplots:acceleration_struct_update} OM/DM update (\fontsize{6.0}{0.0}\selectfont only on TF change)
		\end{tabular}
		};

	\end{tikzpicture}
	
	\vspace{-4mm}
	\caption{Rendering performance for ray marching on the stag beetle dataset with TF6 from~\autoref{fig:TFs_sub6} with viewport size 2160\textsuperscript{2} with early ray termination for different ESS variants. Additionally, occupancy/distance map update times are shown (these only apply if the TF is updated).}
	\label{fig:timing_frame_rendering}
	\vspace{-4mm}
\end{figure}
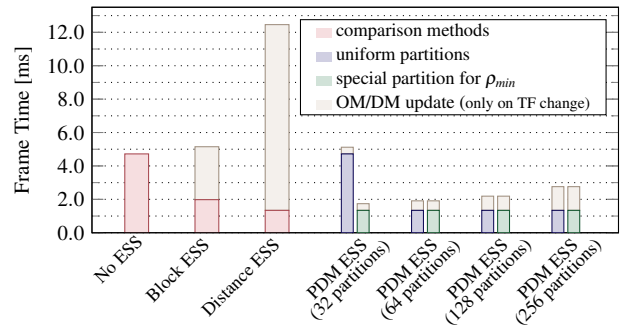

\input{Figure_TFs}
We use a few representative TFs for timing measurements in our experiments (see~\autoref{fig:transfer_functions_used}). The choice typically affects both the runtime of computing $D^\prime$ as well as the actual frame rendering times. Computation time of $D^\prime$ is affected because the number of maps to be combined depends on the TF used. Frame rendering times are affected because depending on the number and configuration of the partitions, both intensity values that influence the rendering as well as those that do not can fall into the same partition. If this happens this may degrade runtime performance since voxels of those intensities will be handled as non-skippable. One occurrence of this issue can be seen in~\autoref{fig:timing_frame_rendering} for the setup with 32 uniform partitions. The required number of partitions to prevent measurable performance degradations depends on the volume and TF used.

In~\autoref{table:timings_distance_maps_computation} we show the computation times for generating and updating the PDMs for a set of partition numbers. We compare the timings with the original distance map approach~\cite{deakin_2019} which has to recompute the occupancy and the distance map on a TF change (we report the combined time). In our partitioned distance map approach we can precompute the POMs and PDMs.
The more partitions are used, the longer it takes to compute $D^\prime$. On the other hand, using fewer partitions may affect frame rendering performance negatively (see thoughts in the beginning of this section and regarding ESS run-time performance in~\autoref{sec:POMs_and_PDMs_explained}). Comparing the computation time of the original distance map, we see an improvement of up to factor 10 for 256 partitions and up to factor 30 for 16 partitions. As one can see the TF used affects the actual timings. On the stag beetle dataset using TF2 (which maps all intensity values to non-transparent output values) our approach performs worse than the original implementation by Deakin and Knackstedt~\cite{deakin_2019}. This is an edge case where the original occupancy map implementation is able to immediately return when computing a block's occupancy state (since all blocks are occupied). In contrast to this, individual POMs may likely have unoccupied blocks depending on the distribution of intensity values in the volume dataset (if all voxels of a block contain intensities from outside the partition's intensity range).
Computing the distance map is especially fast for this edge case since the occupancy map is fully occupied and thus, the distance computation can be exited early.

In~\autoref{fig:timing_frame_rendering_different_datasets_vs_tfs_vs_ess_types} timings for different volume datasets rendered using different TFs are shown. Rendering times in our PDM approach are similar compared to the original distance map approach by Deakin and Knackstedt~\cite{deakin_2019}. For comparison, we also show the timings measured for their block ESS method.
In~\autoref{fig:timing_frame_rendering} timings for rendering the stag beetle dataset using TF6 are compared for different ESS methods. For our method we show timings when using different numbers of intensity partitions, both with uniform partition of the intensity space as well as using a special partition for $\rho_{min}$. For TF6, a different number of intensity partitions does not change rendering times significantly with the exception of 32 partitions where the uniform partitioning breaks down. The reason is that TF6 maps different intensities from the \ordinalnum{1} partition to zero (e.g. $\rho_{min}$) and non-zero alpha values resulting in non-skippable intensity values. The stag beetle dataset contains a high number of voxels with intensity $\rho_{min}$ which degrades ESS performance. With a sufficient number of partitions, all intensities of the first partition are transformed to the same class of mapped alpha values and no performance breakdown occurs. A special partition for $\rho_{min}$ in the 32 partitions case can also prevent the performance degradation.

\begin{figure}[!h]
		\centering
		\begin{tikzpicture}[
			node font=\footnotesize,
			farbe/.style={draw=#1!80!black,fill=#1!20},
			/pgfplots/every axis/.style={
				ybar stacked,
				xtick=data,
				xmin=0.25,
				xmax=6.30,
				ymin=0,
				ymax=3.0,
				ybar=3pt,
				bar width=4.5pt, 
				legend style={nodes={scale=0.8, transform shape}},
				legend image code/.code={%
											\draw[#1, draw=none] (0cm,-0.1cm) rectangle (0.2cm,0.033cm);
				},
			},
			]
			
			\pgfplotsset{every axis title/.append
				style={
					at={(1.045,0.785)},
					rotate=270,
					anchor=east,
					right,
				}
			}
			
			\definecolor{barcolor1_ert}{RGB}{210, 90, 100}
			\definecolor{barcolor1_noert}{RGB}{255, 190, 200}
			\definecolor{barcolor2_ert}{RGB}{21, 15, 110}
			\definecolor{barcolor2_noert}{RGB}{121, 115, 210}
			\definecolor{barcolor3_ert}{RGB}{25, 109, 60}
			\definecolor{barcolor3_noert}{RGB}{125, 209, 160}
			\definecolor{barcolor4_ert}{RGB}{180, 190, 10}
			\definecolor{barcolor4_noert}{RGB}{240, 250, 100}
			
			\pgfplotsset{
					height=2.2cm, width=7cm,
					scale only axis, 
					legend style={draw=none,legend cell align=left,}
			}
			% Barplots
			\begin{axis}[
				name=main_axis,
				bar shift=-18pt,
				title=\textbf{manix dataset},
				xticklabels={TF1, TF2, TF3, TF4, TF5, TF6},
				x tick label style={
					anchor=east,
					%tickwidth = 0,
					%xlabel style={yshift=0.8cm},
				},
				xtick style={draw=none},
				%tickwidth = 0,
				xtick pos=bottom,
				y tick label style={
					/pgf/number format/fixed,
					/pgf/number format/fixed zerofill,
					/pgf/number format/precision=1
        },
				ytick pos=left,
				ytick={0,0.5,...,3},
				minor y tick num={1},
				ylabel={Frame Time [ms]},			
				ymajorgrids=true,
				yminorgrids=true,
				major grid style={dotted,black},
				minor grid style={dotted,black},
				]
				\addplot[farbe=barcolor1_ert] coordinates
				{(1, 0.94) (2, 0.93) (3, 1.70) (4, 1.30) (5, 1.32) (6, 1.15)};
				\label{pgfplots:manix_c1r1}
				\addplot[farbe=barcolor1_noert] coordinates
				{(1, 0.75) (2, 0.76) (3, 0.0) (4, 0.39) (5, 0.40) (6, 0.57)};
			\end{axis}
			
			\begin{axis}[bar shift=-12pt,hide axis]
				\addplot+[farbe=barcolor2_ert] coordinates
				{(1, 0.61) (2, 0.94) (3, 0.94) (4, 0.71) (5, 0.86) (6, 0.73)};
				\label{pgfplots:manix_c2r1}
				\addplot+[farbe=barcolor2_noert] coordinates
				{(1, 0.81) (2, 0.78) (3, 0.01) (4, 0.51) (5, 0.47) (6, 0.65)};
			\end{axis}
					
			\begin{axis}[bar shift=-6pt,hide axis]
				\addplot+[farbe=barcolor3_ert] coordinates
				{(1, 0.51) (2, 0.94) (3, 0.73) (4, 0.56) (5, 0.72) (6, 0.62)};
				\label{pgfplots:manix_c3r1}
				\addplot+[farbe=barcolor3_noert] coordinates
				{(1, 0.79) (2, 0.79) (3, 0.01) (4, 0.49) (5, 0.45) (6, 0.62)};
			\end{axis}
			
			\begin{axis}[bar shift=0pt,hide axis]
				\addplot+[farbe=barcolor4_ert] coordinates
				{(1, 0.68) (2, 0.94) (3, 0.74) (4, 0.58) (5, 0.73) (6, 0.62)};
				\label{pgfplots:manix_c4r1}
				\addplot+[farbe=barcolor4_noert] coordinates
				{(1, 0.78) (2, 0.76) (3, 0.01) (4, 0.49) (5, 0.45) (6, 0.62)};
			\end{axis}
			
			\node[draw,fill=white,inner sep=0pt,below left=0.5em] at (main_axis.north east) {
			\small

			\begin{tabular}{llllll}
			\ref{pgfplots:manix_c1r1} no ESS & \ref{pgfplots:manix_c3r1} standard distance map ESS~\cite{deakin_2019}\\
			\ref{pgfplots:manix_c2r1} block ESS~\cite{deakin_2019} & \ref{pgfplots:manix_c4r1} PDM ESS (64 partitions)
			\end{tabular}
			};
			
		\end{tikzpicture}
		\label{fig:subfig_frame_rendering_timings_tfs_2_ess_types_manix_dataset}

		\vspace{-3.5mm}

		\centering
		\begin{tikzpicture}[
			node font=\footnotesize,
			farbe/.style={draw=#1!80!black,fill=#1!20},
			/pgfplots/every axis/.style={
				ybar stacked,
				xtick=data,
				xmin=0.25,
				xmax=6.30,
				ymin=0,
				ymax=5.5,
				ybar=3pt,
				bar width=4.5pt,
				legend style={nodes={scale=0.8, transform shape}},
				legend image code/.code={%
											\draw[#1, draw=none] (0cm,-0.1cm) rectangle (0.2cm,0.033cm);
				},
			},
			]
			
			\pgfplotsset{every axis title/.append
				style={
					at={(1.04,0.98)},
					rotate=270,
					anchor=east,
					right,
				}
			}
			
			\definecolor{barcolor1_ert}{RGB}{210, 90, 100}
			\definecolor{barcolor1_noert}{RGB}{255, 190, 200}
			\definecolor{barcolor2_ert}{RGB}{21, 15, 110}
			\definecolor{barcolor2_noert}{RGB}{121, 115, 210}
			\definecolor{barcolor3_ert}{RGB}{25, 109, 60}
			\definecolor{barcolor3_noert}{RGB}{125, 209, 160}
			\definecolor{barcolor4_ert}{RGB}{180, 190, 10}
			\definecolor{barcolor4_noert}{RGB}{240, 250, 100}
			
			\pgfplotsset{
					height=2.0cm, width=7cm,
					scale only axis, 
					legend style={draw=none,legend cell align=left,}
			}
			% Barplots
			\begin{axis}[
				name=main_axis,
				bar shift=-18pt,
				title=\textbf{stag beetle dataset},
				xticklabels={TF1, TF2, TF3, TF4, TF5, TF6},
				x tick label style={
					anchor=east,
					%tickwidth = 0,
					%xlabel style={yshift=0.8cm},
				},
				xtick style={draw=none},
				%tickwidth = 0,
				xtick pos=bottom,
				y tick label style={
					/pgf/number format/fixed,
					/pgf/number format/fixed zerofill,
					/pgf/number format/precision=1
        },
				ytick pos=left,
				ytick={0,1,...,5.5},
				minor y tick num={1},
				ylabel={Frame Time [ms]},			
				ymajorgrids=true,
				yminorgrids=true,
				major grid style={dotted,black},
				minor grid style={dotted,black},
				]

				\addplot[farbe=barcolor1_ert] coordinates
				{(1, 4.61) (2, 4.56) (3, 5.15) (4, 4.84) (5, 5.10) (6, 4.82)};
				\label{pgfplots:stagbeetle_c1r1}
				\addplot[farbe=barcolor1_noert] coordinates
				{(1, 0.45) (2, 0.57) (3, 0.0) (4, 0.27) (5, 0.01) (6, 0.29)};
			\end{axis}
			
			\begin{axis}[bar shift=-12pt,hide axis]
				\addplot+[farbe=barcolor2_ert] coordinates
				{(1, 1.82) (2, 4.62) (3, 1.57) (4, 1.99) (5, 2.41) (6, 2.00)};
				\label{pgfplots:stagbeetle_c2r1}
				\addplot+[farbe=barcolor2_noert] coordinates
				{(1, 0.66) (2, 0.52) (3, 0.00) (4, 0.44) (5, 0.06) (6, 0.50)};
			\end{axis}
					
			\begin{axis}[bar shift=-6pt,hide axis]
				\addplot+[farbe=barcolor3_ert] coordinates
				{(1, 1.19) (2, 4.59) (3, 0.68) (4, 1.34) (5, 1.74) (6, 1.34)};
				\label{pgfplots:stagbeetle_c3r1}
				\addplot+[farbe=barcolor3_noert] coordinates
				{(1, 0.64) (2, 0.54) (3, 0.00) (4, 0.42) (5, 0.05) (6, 0.48)};
			\end{axis}
			
			\begin{axis}[bar shift=0pt,hide axis]
				\addplot+[farbe=barcolor4_ert] coordinates
				{(1, 1.19) (2, 4.61) (3, 0.69) (4, 1.34) (5, 1.74) (6, 1.34)};
				\label{pgfplots:stagbeetle_c4r1}
				\addplot+[farbe=barcolor4_noert] coordinates
				{(1, 0.63) (2, 0.51) (3, 0.00) (4, 0.42) (5, 0.05) (6, 0.48)};
			\end{axis}
			
		\end{tikzpicture}
		\label{fig:subfig_frame_rendering_timings_tfs_2_ess_types_stag_beetle_dataset}

		\vspace{-3.5mm}

		\centering
		\begin{tikzpicture}[
			node font=\footnotesize,
			farbe/.style={draw=#1!80!black,fill=#1!20},
			/pgfplots/every axis/.style={
				ybar stacked,
				xtick=data,
				xmin=0.25,
				xmax=6.30,
				ymin=0,
				ymax=6.5,
				ybar=3pt,
				bar width=4.5pt,
				legend style={nodes={scale=0.8, transform shape}},
				legend image code/.code={%
											\draw[#1, draw=none] (0cm,-0.1cm) rectangle (0.2cm,0.033cm);
				},
			},
			]
			
			\pgfplotsset{every axis title/.append
				style={
					at={(1.04,0.975)},
					rotate=270,
					anchor=east,
					right,
				}
			}
			
			\definecolor{barcolor1_ert}{RGB}{210, 90, 100}
			\definecolor{barcolor1_noert}{RGB}{255, 190, 200}
			\definecolor{barcolor2_ert}{RGB}{21, 15, 110}
			\definecolor{barcolor2_noert}{RGB}{121, 115, 210}
			\definecolor{barcolor3_ert}{RGB}{25, 109, 60}
			\definecolor{barcolor3_noert}{RGB}{125, 209, 160}
			\definecolor{barcolor4_ert}{RGB}{180, 190, 10}
			\definecolor{barcolor4_noert}{RGB}{240, 250, 100}

			\pgfplotsset{
					height=2.0cm, width=7cm,
					scale only axis, 
					legend style={draw=none,legend cell align=left,}
			}
			% Barplots
			\begin{axis}[
				name=main_axis,
				bar shift=-18pt,
				title=\textbf{kingsnake dataset},
				xticklabels={TF1, TF2, TF3, TF4, TF5, TF6},
				x tick label style={
					anchor=east,
					%tickwidth = 0,
					%xlabel style={yshift=0.8cm},
				},
				xtick style={draw=none},
				%tickwidth = 0,
				xtick pos=bottom,
				y tick label style={
					/pgf/number format/fixed,
					/pgf/number format/fixed zerofill,
					/pgf/number format/precision=1
        },
				ytick pos=left,
				ytick={0,1,...,6.5},
				minor y tick num={1},
				ylabel={Frame Time [ms]},			
				ymajorgrids=true,
				yminorgrids=true,
				major grid style={dotted,black},
				minor grid style={dotted,black},
				]

				\addplot[farbe=barcolor1_ert] coordinates
				{(1, 1.79) (2, 1.71) (3, 6.14) (4, 2.93) (5, 5.77) (6, 2.63)};
				\label{pgfplots:kingsnake_c1r1}
				\addplot[farbe=barcolor1_noert] coordinates
				{(1, 4.31) (2, 4.38) (3, 0.0) (4, 3.18) (5, 0.35) (6, 3.51)};
			\end{axis}
			
			\begin{axis}[bar shift=-12pt,hide axis]
				\addplot+[farbe=barcolor2_ert] coordinates
				{(1, 1.39) (2, 1.73) (3, 3.42) (4, 1.45) (5, 4.83) (6, 2.35)};
				\label{pgfplots:kingsnake_c2r1}
				\addplot+[farbe=barcolor2_noert] coordinates
				{(1, 4.63) (2, 4.41) (3, 0.01) (4, 4.00) (5, 0.66) (6, 3.81)};
			\end{axis}
					
			\begin{axis}[bar shift=-6pt,hide axis]
				\addplot+[farbe=barcolor3_ert] coordinates
				{(1, 1.27) (2, 1.72) (3, 2.64) (4, 1.15) (5, 4.39) (6, 2.25)};
				\label{pgfplots:kingsnake_c3r1}
				\addplot+[farbe=barcolor3_noert] coordinates
				{(1, 4.75) (2, 4.40) (3, 0.00) (4, 3.73) (5, 0.51) (6, 3.77)};
			\end{axis}
			
			\begin{axis}[bar shift=0pt,hide axis]
				\addplot+[farbe=barcolor4_ert] coordinates
				{(1, 1.26) (2, 1.72) (3, 2.64) (4, 1.15) (5, 4.39) (6, 2.25)};
				\label{pgfplots:kingsnake_c4r1}
				\addplot+[farbe=barcolor4_noert] coordinates
				{(1, 4.71) (2, 4.40) (3, 0.00) (4, 3.72) (5, 0.52) (6, 3.76)};
			\end{axis}
			
		\end{tikzpicture}
		\label{fig:subfig_frame_rendering_timings_tfs_2_ess_types_kingsnake_dataset}
		
	\vspace{-5mm}
	\caption{Frame rendering times for different datasets (viewport size 2160\textsuperscript{2}) with different TFs and ESS variants. Stacked bars indicate rendering time w/o early ray termination (darker/lighter). PDM ESS setup: 64 uniform intensity partitions for TF2-TF6, for TF1: special partition for intensity value $\rho_{min}$ and the other intensity values uniformly partitioned into the remaining 63 partitions.}

	\label{fig:timing_frame_rendering_different_datasets_vs_tfs_vs_ess_types}
\end{figure}
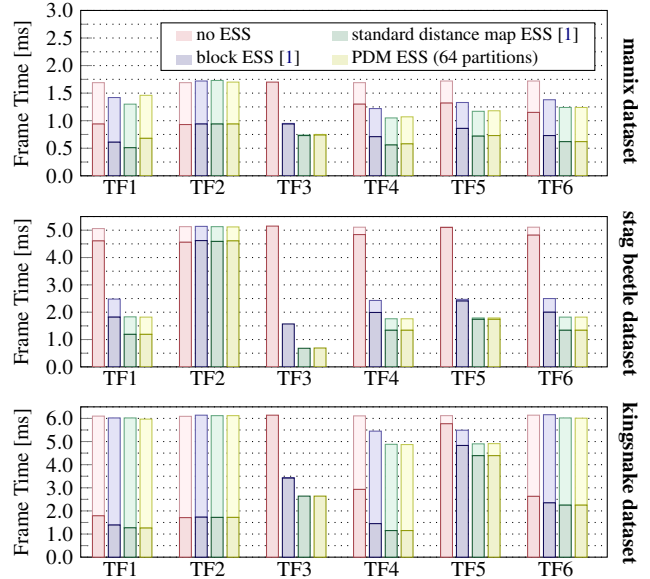

\section{Discussion}
Our method allows for TF updates with low performance impact, nevertheless update times are higher than in the method by Faludi et al.~\cite{faludi_2021} which does not require recomputing the acceleration structure.
On the other hand, our method is faster in terms of rendering performance as it extends the fastest currently known volume rendering method (distance-map based ESS) allowing frequent TF updates while maintaining higher frame rates and reducing latency.

Our approach requires additional GPU memory for storing the PDMs. The total extra amount of memory required depends on the block size $b$ of the reduced volume and the number of partitions $n$. Since the distance maps are textures of unsigned 8-bit integer data type, the required memory amount in bytes is $n \times \frac{volumesize}{b^3}$ with $volumesize$ being the number of voxels in the original volume dataset. The POMs, which are needed to compute the PDMs, are just an intermediate result. Each partition is computed independently from each other, therefore only one additional texture for computing the individual POMs is needed. For this purpose, one of the flip-flopping textures required for computing $D'$ can be reused.

\section{Conclusion}
We improved distance map based ESS in volume rendering to better support frequent TF updates.
Our method is valuable when volume rendering performance is the highest priority but also frequent TF updates need to be done. Nevertheless changing the TF will impact performance more than in the method by Faludi et al.~\cite{faludi_2021}. Drawbacks of our approach are additional initialization time for creating the PDMs as well as extra GPU memory needed which does affect the maximum possible volume size for in-core volume rendering.

%% if specified like this the section will be committed in review mode
\acknowledgments{This work was funded as part of the RTI-strategy Lower Austria 2027.}

\bibliographystyle{abbrv-doi-hyperref-narrow}

\bibliography{references}
\end{document}